\documentclass[12pt]{iopart}

\usepackage{cite}
\usepackage{graphicx}
\usepackage{bm}
\usepackage{color}

\usepackage{xcolor}
\usepackage[colorlinks = true,
            linkcolor = blue,
            urlcolor  = blue,
            citecolor = blue,
            anchorcolor = blue]{hyperref}

\begin{document}
\title[]{Transport of relativistic jet in the magnetized intergalactic medium}

\author{W. P. Yao$^{1}$, B. Qiao$^{1,2,3\ast}$, Z. Xu$^{1}$, H. Zhang$^{3}$, Z. H. Zhao$^{1}$, H. X. Chang$^1$, C. T. Zhou$^{1,3}$, S. P. Zhu$^{3,4}$, and X. T. He$^{1,3}$}
\address{$^1$ Center for Applied Physics and Technology, HEDPS, and State Key Laboratory of Nuclear Physics and Technology, School of Physics, Peking University, Beijing 100871, P. R. China}
\address{$^2$ Collaborative Innovation Center of IFSA (CICIFSA), Shanghai Jiao Tong University, Shanghai 200240, China}
\address{$^3$ Institute of Applied Physics and Computational Mathematics, Beijing 100094, P. R. China}
\address{$^4$ Graduate School of China Academy of Engineering Physics, P. O. Box 2101, Beijing 100088, China}

\ead{bqiao@pku.edu.cn}

\begin{abstract}

The kinetic effects of magnetic fields on the transport of relativistic jet in the intergalactic medium remain uncertain, especially for their perpendicular component.
By particle-in-cell simulations, we find that when only jet electrons are fully magnetized, they are directly deflected by magnetic field, but jet protons are mainly dragged by collective charge-separation electric field.
However, when both electrons and protons are fully magnetized, the contrary is the case.
Their balance tremendously distorts the jet density and electromagnetic fields, leading to enormous energy exchange between different species and fields.
As a result, the electron spectrum energy distribution (SED) gets reshaped and the power law slope of the SED decreases as the magnetic field strength increases. 
In other words, we may infer that magnetic fields around the relativistic jet play a crucial role in shaping the observed SED.

\end{abstract}

\newpage 

\section{INTRODUCTION}

Astrophysical jets are collimated outflow which are commonly observed in the Universe \cite{Vietri2003, Fabian2012}.
Associated with the most exciting high-energy astrophysical phenomena, e.g. Gamma-ray burst (GRB) \cite{Bykov2011, Kumar2015} and tidal disruption events \cite{Piran2015, zzq2017}, astrophysical jets are ejected from an astronomical object, or a central engine \cite{Bridle1984}, such as supermassive black holes \cite{Young1991} and other compact objects \cite{Migliari2006, Pavan2014}. 
Among them, relativistic jets are perhaps the most energetic ones \cite{Kynoch2018}, such as blazar jets from active galactic nucleis (AGNs) \cite{Giannios2009, Sironi2014}.
Study of the relativistic jets transport in the intergalactic medium (IGM) can provide us with a better understanding of their host sources and surrounding environments.

The first recorded jet observation is in {\it M87} at 1918 \cite{curtis1918descriptions}. 
After a century of dedicated studies, scientists have collected abundant data via state-of-art space telescopes, such as Very Large Array \cite{marschern2010jets} and Very Long Baseline interferometry (VLBI) \cite{Hovatta2016}. 
Although no consensus has been reached concerning the basic parameters of relativistic jets (e.g. the component, the Mach number, the jet-to-IGM density ratio and the magnetic fields around) \cite{2010fmja.conf}, some characteristic features of them are well accepted. 
Perhaps the most important one lies in their non-thermal radiation -- flat spectrum energy distributions (SED) are often observed from the X-ray of luminous relativistic jet \cite{Tavecchio2008, Chen2018}. 

Of all those basic parameters, magnetic fields play a critical role in the jet-IGM interaction \cite{Zamaninasab2014, Bruni:2017xgh}.
Observationally, the orientation of magnetic field is closely related to the jet sidedness \cite{Laing2006yv}; 
while theoretically, it leads to mode coupling of different instabilities \cite{Tautz2006, Matsumoto2017} and alters the particle acceleration \cite{Dieckmann2010, Bykov2012, Sironi2015, Liu2017}.
For example, as for the parallel component, it suppresses the growth rate of Weibel instability (WBI) \cite{Silva2002, Stockem2008}, reduces the saturation level \cite{Grassi2017}, and prevents the collisionless shock formation \cite{Stockem2007,Bret2016}; 
while as for the perpendicular one, ultra-strong magnetic field supports the relativistic jet transport in a relay manner by quantum electrodynamics effects \cite{yao2017}.

In fact, the formation of a particular jet SED has always been an open question that puzzles both astrophysicist \cite{icke1992collimation, meier2001magnetohydrodynamic} and experimental scientists \cite{gregory2009colliding, Kuramitsu2009}. 
For the former, as observations crucially depend on the viewing angle, it is not enough to narrow down jet parameters \cite{Massaglia2003}. 
While for the latter, it is extremely challenging to generate relativistic jets in laboratory \cite{yuan2015modeling, li2016scaled, yuan2018laboratory}.
More importantly, collective and kinetic effects, such as electric field generation by charge-separation \cite{Tzoufras2006}, as well as magnetic field generation by WBI \cite {Huntington2015} and Kelvin-Helmholtz instability (KHI) \cite{Alves2012}, become crucial for the energy transport, especially in regions like the X-point of magnetic reconnection \cite{Xu2017} and the discontinuity surface of collisionless shock \cite{Marcowith2016}.
Thus, kinetic treatment is necessary to shed new light on the above open questions.

As there are extensive investigations on the parallel component of magnetic field, in this paper, we study the kinetic effect of perpendicular magnetic field on the transport of relativistic jet in the IGM. 
Through two-dimensional (2D) Particle-in-Cell (PIC) simulations, we find that the magnetic field strength (or magnetization rate) plays a crucial role in shaping the jet electron SED.
Specifically speaking, when only jet electrons are fully magnetized, they are directly deflected by magnetic field, but jet protons are indirectly dragged by electrons via collective charge-separation field.
While when both jet electrons and protons are fully magnetized, the contrary is the case.
Under the balance between the magnetic field deflection and charge-separation field drag, the jet density and electromagnetic (EM) fields are distorted tremendously in both cases.
Furthermore, as energy exchange between species becomes the violent, especially for the energetic ones, the SED of electrons gets reshaped.

\section{SIMULATION SETUP}
\label{simsetup}

To self-consistently study the kinetic effects of magnetic fields on the transport of relativistic jet in the IGM, a series of 2D PIC simulations are performed, which capture the fundamental interplay between charged particles and EM fields from first principles.

\begin{figure*}
	\begin{center}
	\includegraphics{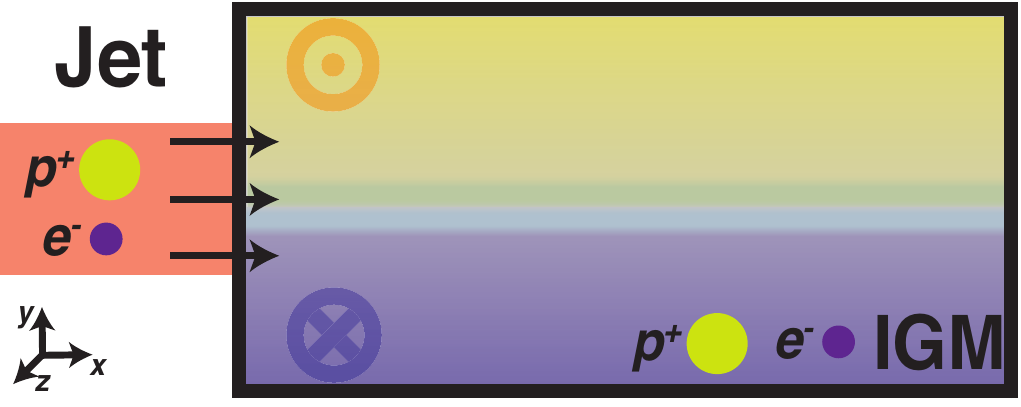}
 	\caption{(color online) Schematic for simulation setup. Relativistic jet (red) is injected from the left boundary, which has a supergaussian transverse profile. Magnetic field is directed towards $+z$ in $y > 0$ region (yellow) and $-z$ in $y < 0$ region (blue).}
	\label{setup}
	\end{center}
\end{figure*}

The simulation setup is shown in Fig.\ref{setup}, both the relativistic jet and the IGM are composed of proton and electron. 
A reduced proton mass $m_i = 100m_e$ ($m_e$ is the electron mass) is used \cite{Bret2010}.
Each specie is represented by 8 marco-particles per cell.
The IGM plasma species are uniformly distributed and follow the Maxwell distribution $f_{p,s} \propto \exp \left( m_sc^2/T_{p,s0} \right)$ ($p$ for IGM plasma and $s=i,e$ for proton and electron), where the initial temperatures are $T_{pe0} = 10$ keV and $T_{pi0} = 1$ keV, which are in the same order of the temperature inferred for AGN jets \cite{Homan2006, Ping2017}.
While the relativistic jet follows a drifted Maxwell-J\"{u}ttner distribution \cite{Melzani2013, Swisdak2013}
\begin{center}
\begin{equation}
	f_{j,s}(\bm{p}) \propto \exp \left[-\frac{\gamma_d}{T_{j,s0}}(\sqrt{1 + \bm{p}^2} - V_d p_x)\right]
\end{equation}
\end{center}
($j$ for relativistic jet) where the initial temperature $T_{je0} = 1$ keV and $T_{ji0} = 0.1$ keV are taken. 
$\bm{p}$ is the momentum and $V_d$ is the drift velocity. 
The Lorentz factor is $\gamma_d = (1-(V_d/c)^2)^{-1/2} = 16$, just like the relativistic baryonic fireball model in GRBs, where most of the initial energy is the bulk kinetic energy of protons \cite{PIRAN1999575}.
The Sobol method \cite{Zenitani2015} is used to load relativistic particles at each step.
Note that the relativistic jet has a supergaussian transverse profile with a width $W_j=10\lambda_{ji}$, so that we can have a global picture of the jet transport, which is different from the width-unlimited model in previous studies \cite{Ardaneh2016, Yao2018}.
Here, $\lambda_{ji} = c/\omega_{ji}$ is the jet proton skin depth, where $c$ is the light speed, $\omega_{ji} = [n_{ji} q_i^2/(\varepsilon_0 m_i \gamma_d)]^{1/2}$ is the plasma frequency of jet proton ($n_{ji}$, and $\varepsilon_0$ are the number density and the vacuum permittivity).
To minimize the influence of the injection boundary, the IGM number density is twice of the jet number density.

The simulation box size is $240\lambda_{ji}$ long and wide, which is divided into $4800 \times 4800$ cells along $x$ and $y$, with the Debye length $\lambda_{De} = [\varepsilon_0kT_{pe0}/(n_{pe} q_e^2)]^{1/2}$ resolved ($n_{pe}$ is the IGM electron density and $k$ is the Boltzmann constant).
Additionally, to avoid the numerical Cherenkov instability \cite{Vay2014} caused by the relativistic speed, the 3rd-order B-Spline shape function, 5th-order weighting, and 6th-order finite different scheme for solving Maxwell equations have been adopted.

While for the magnetic fields, motivated by the ``striped wind model'' in Pulsar Wind Nebulae \cite{Bogovalov1999, Lyubarsky2003} and the recent VLBI observation \cite{Gomez2016}, we consider the effects of toroidal component of magnetic field that surround the relativistic jet.
For simplicity, this external applied magnetic field is directed towards $z$-axis and distributed non-uniformly along $y$-axis, which can be expressed as $B_{z0} = B_0 \tanh{(y/W_b)}$, where $-120\lambda_{ji} < y < 120\lambda_{ji}$.
The magnetization rate is defined as $\sigma = \Omega_{je} / \omega_{je}$, where $\Omega_{je} = q_e B_0 / (\gamma_d m_e)$ is the cyclotron frequency of jet electron and $\omega_{je} = [n_{je} q_e^2/(\varepsilon_0 m_e \gamma_d)]^{1/2}$ is the plasma frequency of jet electron.
In our presented cases, we choose $W_b = 5\lambda_{ji}$.
Parametric study shows that our results are insensitive to $W_b$.

In the following sections, three typical cases are performed:

\begin{enumerate}

\item $\sigma = 0$ case for the unmagnetized jet-IGM interaction.

In this case, a global view of relativistic jet transport in the IGM is given, with the characteristic feature of the WBI and KHI at the jet body and lateral, respectively.

\item $\sigma = 2$ case for the weakly magnetized jet-IGM interaction -- jet electrons are fully magnetized ($\Omega_{je} = 2\omega_{je}$), while jet protons are not ($\Omega_{ji} = 0.2 \omega_{ji}$).

In this case, the transport of jet electrons is directly deflected by the external applied magnetic field (or, $B_{z0}$), while that of jet protons is dominated by the electric field, which is collectively induced by charge-separation.

\item $\sigma = 10$ case, both jet species are fully magnetized ($\Omega_{je} = 10\omega_{je}$ and $\Omega_{ji} =  \omega_{ji}$).

In this case, the transport of jet protons is deflected directly by $B_{z0}$, while that of jet electrons is dominated by charge-separation electric field.

\end{enumerate}

\section{UNMAGNETIZED JET-IGM INTERACTION}

\noindent
Before we dive into the $B_{z0}$ effects on the jet-IGM interaction, let's take a look at the full picture of the unmagnetized jet-IGM interaction ($\sigma = 0$ case) in this section.

\begin{figure*}
	\begin{center}
	\includegraphics{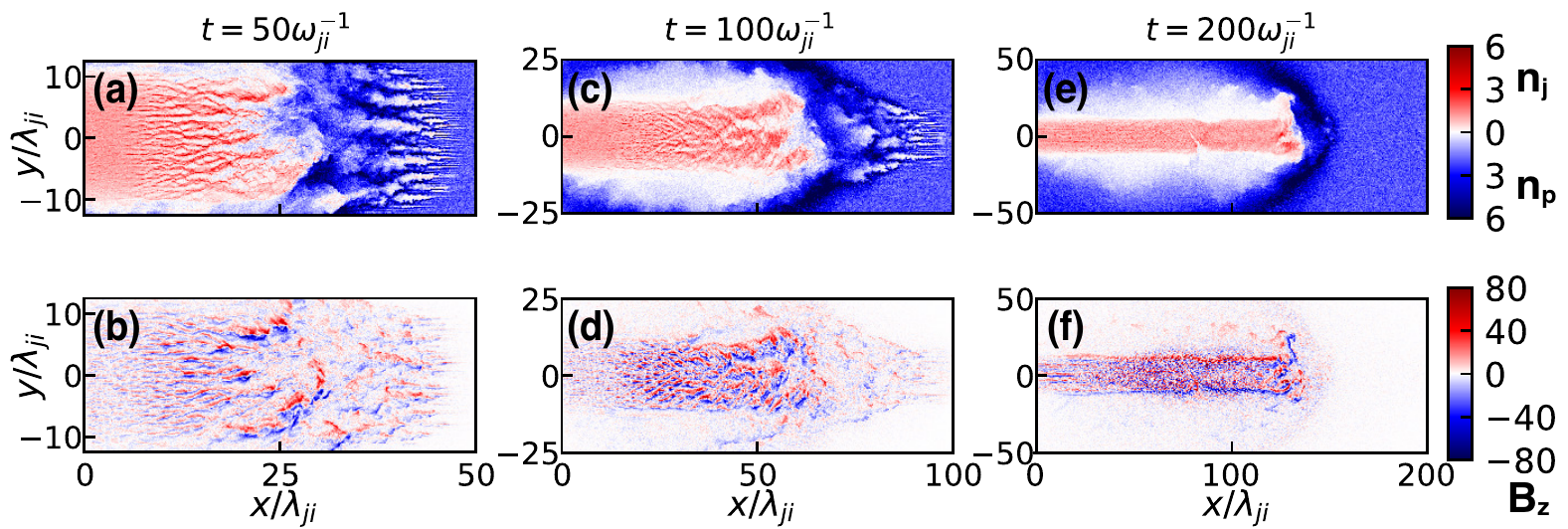}
 	\caption{(color online) Global evolution of the unmagnetized jet-IGM interaction at $t=50/100/200\omega_{ji}^{-1}$. The upper row is the electron number density of jet (red, $n_j$) and the IGM (blue, $n_p$), both normalized by the initial jet number density $n_0$. While the lower row is the self-generated $B_z$, normed by $B_0 = m_e \omega_{je} / q_e$. The proton number density is not shown here because it is almost the same with the electron's.}
	\label{global}
	\end{center}
\end{figure*}

Firstly, Fig.\ref{global} demonstrates the global evolution of the relativistic jet transport in the unmagnetized IGM.
At the beginning ($t=50\omega_{ji}^{-1}$), the anisotropy induced by the jet particles excites the well-known WBI, which results in filaments of both densities (jet and the IGM) and self-generated $B_z$ in Fig.\ref{global} (a) and (b). 
Later ($t=100\omega_{ji}^{-1}$), the WBI is getting saturated, as the filaments merge into thicker ones in Fig.\ref{global} (c) and (d). 
Note that in this width-limited model (or, ``global model" \cite{Nishikawa2016}), the IGM density is pushed aside by the jet pressure due to the transverse expansion effects.
At last ($t=200\omega_{ji}^{-1}$), a bow-shock structure is formed in Fig.\ref{global} (e), and a DC magnetic field structure \cite{Grismayer2013} is generated at the lateral of the relativistic jet by KHI in Fig.\ref{global} (f).

\begin{figure*}
	\begin{center}
	\includegraphics{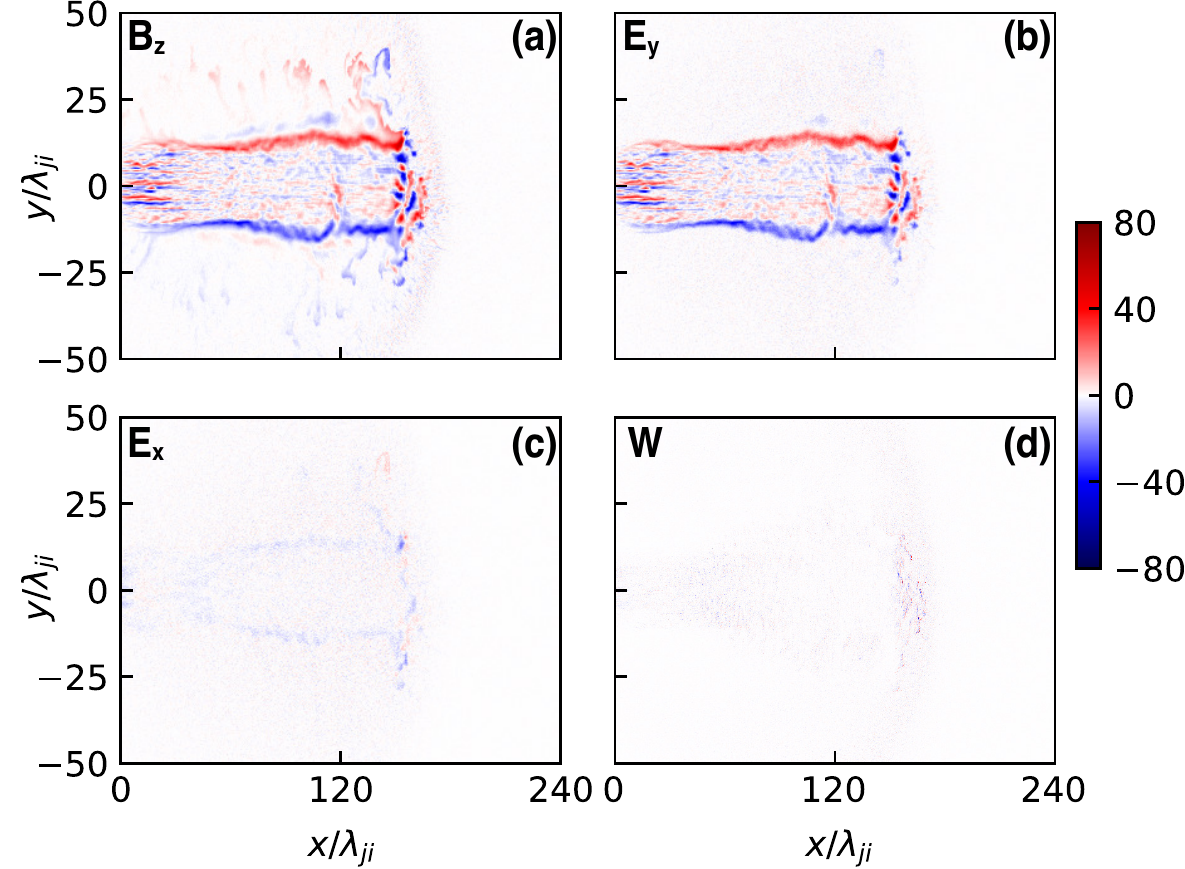}
 	\caption{(color online) EM fields ($B_z$, $E_y$ and $E_x$) and the work (done by the Lorentz force $W$) distributions at $t = 240\omega_{ji}^{-1}$ in $\sigma = 0$ case. $B_z$ is normed by $B_0 = m_e \omega_{je} / q_e$, $E_y$ and $E_x$ are by $E_0 = m_e \omega_{je} c / q_e$ and $W$ is by $W_0 = m_e \omega_{je} c^2$.}
	\label{field}
	\end{center}
\end{figure*}

Secondly, the EM fields and the work (done by the Lorentz force) distributions at the end of the transport ($t=240\omega_{ji}^{-1}$) are shown in Fig.\ref{field}.
Because WBI dominates, on the one hand, the transverse EM fields ($B_z$ and $E_y$) is much larger than the longitudinal one ($E_x$);
on the other hand, as $\beta_{x,s} = V_d / c \approx 1$ ($s = ji, je$) and $\bm{\beta_s} \times \bm{B} = \bm{E}$, we get $E_y \approx B_z$, just as Fig.\ref{field} (a) and (b) show.
Note that far from the jet body ($25 \lambda_{ji} < |y| < 50 \lambda_{ji}$), $E_y \neq B_z$, because $\beta_{x,s} < 1$ there after modulated by the KHI.
Accrodingly, almost no work has been done by the Lorentz force ($W = \bm{v} \cdot \bm{F} = q_s \bm{v} \cdot (\bm{E} + \bm{v} \times \bm{B}/c)$), as is shown in Fig.\ref{field} (d), which means few energy is exchanged between particles and fields in $\sigma = 0$ case, and energy transfer is mainly restricted between different species.

\begin{figure*}
	\begin{center}
	\includegraphics{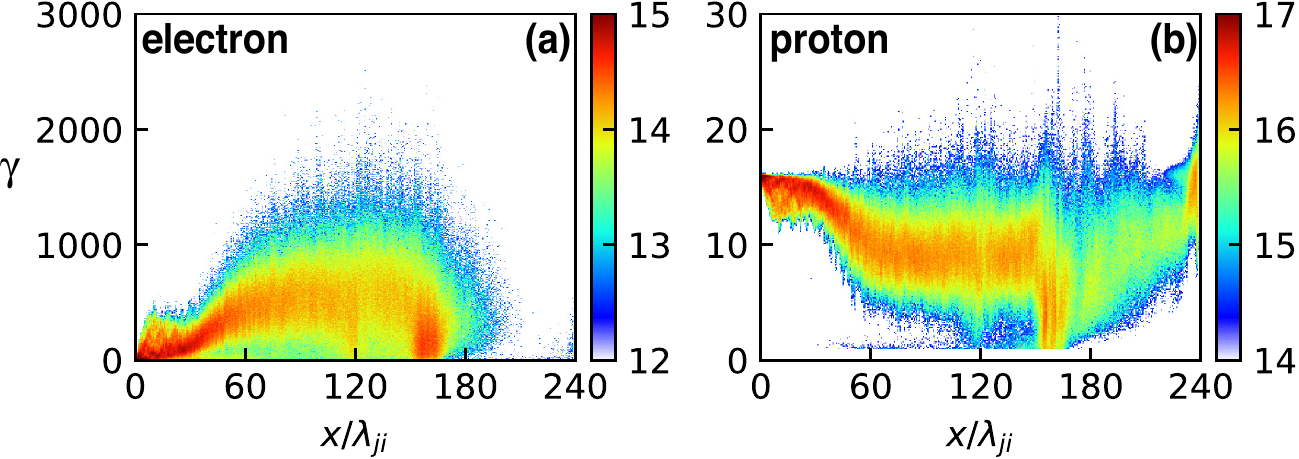}
 	\caption{(color online) Energy spatial distribution of jet electron (a) and proton (b) between $|y| < 20\lambda_{ji}$ at $t=240\omega_{ji}^{-1}$ in $\sigma = 0$ case.  The colorbar represents the particle number in logarithm.}
	\label{energy}
	\end{center}
\end{figure*}

Thirdly, let's check the particle dynamics in $\sigma = 0$ case.
Figure \ref{energy} shows the energy spatial distributions of jet particles.
At the jet injection region ($0 < x < 30 \lambda_{ji}$), the WBI is at its linear stage, so the electron energy increases a few from $\gamma = 16$ to about $400$.
Accordingly, the proton energy decreases that much from $\gamma = 16$ to about $12$.
While at the jet body region ($30\lambda_{ji} < x < 150 \lambda_{ji}$), the WBI gradually grows into its nonlinear stage and more energy has been transferred.
Moreover, owing to the few work done by the Lorentz force, few protons get accelerated.
Note that between $180\lambda_{ji} < x < 240\lambda_{ji}$, the jet charge separation there will be balanced by the IGM plasmas.
%Note that between $180\lambda_{ji} < x < 240\lambda_{ji}$, there are some jet protons there. 
%They come from the leading edge of the jet, whose density is initially rising from $0$ to $n_j$.
%However, almost no electrons are left there, because they are either scattered away transversely or trapped around the jet head region by EM fields.
%The jet charge separation here will be balanced by the IGM plasmas.  

In short, in $\sigma = 0$ case, the jet body will be dominated by the WBI and the jet lateral by the KHI, resulting in a bow shock structure and a DC magnetic field structure, respectively.
Besides, transverse EM fields will be much larger than the longitudinal one and the Lorentz force only does few work.
Moreover, energy is mainly transferred from proton to electron around the jet body via kinetic instabilities.

\section{MAGNETIZED JET-IGM INTERACTION}

With the full picture of the unmagnetized jet-IGM interaction in mind, now let's dive into the $B_{z0}$ effects on the relativistic jet transport from two different regimes -- weakly and strongly magnetized cases.
As $\sigma = \Omega_{je}/\omega_{je}$ and $\Omega_{ji} / \omega_{ji} = (m_e/m_i)^{1/2} \sigma = \sigma/10$, weakly magnetized regime is defined when $1 \le \sigma < 10$ (or $\Omega_{je} \ge \omega_{je}$ but $\Omega_{ji} < \omega_{ji}$), in which electrons are fully magnetized but protons are not; while strongly one is defined when $\sigma \ge 10$ (or $\Omega_{je} > \omega_{je}$ and $\Omega_{ji} \ge \omega_{ji}$), in which both are fully magnetized.

\subsection{Weakly magnetized regime (for the presented case, $\sigma = 2$)}

In fact, the cyclotron radius of each charged particles can be written by 
\begin{center}
\begin{equation}
	r_s = \frac{\gamma_s m_s \beta_s c}{q_s B_{z0}} = \frac{\gamma_s \beta_s \alpha_s^{1/2}}{\sigma} \lambda_s
	\label{radius}
\end{equation}
\end{center}
where $\alpha_s = m_s / m_e$ ($s = ji, je$). 
In $\sigma = 2$ case, the cyclotron radius of jet protons is $r_{ji} = 80 \lambda_{ji}$, that of jet electrons is $r_{je} = 0.8 \lambda_{ji}$, while that of the IGM specie is about $v_{th,s}/(\gamma_d c)$ times smaller, where $v_{th,s} = (kT_s/m_s)^{1/2}$ ($s = pi, pe$) is the thermal velocity.
Besides the difference in radius, the cyclotron direction of electrons and protons is opposite.
%Specifically speaking, as is shown in Fig.\ref{setup}, at the upper half of our simulation box, $B_{z0}$ is in $+z$ direction, with relativistic jet injecting towards $+x$ direction, the electrons are deflected towards $+y$ direction and the ions are deflected towards $-y$ direction. 
%While at the lower half of our simulation box, the deflection direction for each species is the opposite as $B_{z0}$ is in $-z$ direction.
Thus instinctively, the transport of jet electrons should have been diverged a lot, that of jet protons been converged a little, and that of IGM particles been trapped.

However, when considering the collective plasma effects, namely, the charge-separation electric field (owing to the different response of electron and proton to the magnetic field)
and the induction electric field (due to the magnetic field gradient $\partial_t \bm{E} = \bm{\nabla} \times \bm{B} - \mu_0 \bm{J}$), the density distributions of electron and proton remain the same, but the EM fields and energy transport have been tremendously altered.

\begin{figure*}
	\begin{center}
	\includegraphics{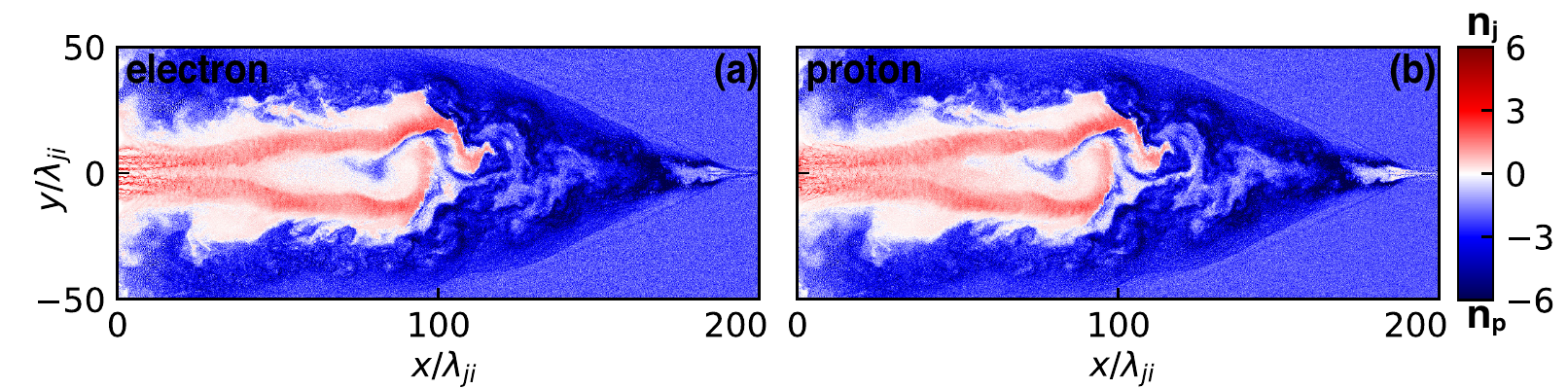}
 	\caption{(color online) Density distributions of both electron (a) and proton (b) at $t=200\omega_{ji}^{-1}$ for in $\sigma = 2$ case. Jet densities are in red $n_j$, while the IGM densities are in blue $n_p$, both normalized by the initial jet number density $n_0$.}
	\label{dens}
	\end{center}
\end{figure*}

To begin with, Fig.\ref{dens} shows the electron and proton density distributions at $t=200\omega_{ji}^{-1}$ for $\sigma = 2$ case.
Jet electrons are fully magnetized with $\Omega_{je} = 2 \omega_{je}$, so their transport is directly diverged by magnetic fields, but with a much larger radius dragged by jet protons through the charge-separation electric field.
However, jet protons are only slightly magnetized with $\Omega_{ji} = 0.2 \omega_{ji}$, so their transport is mainly dragged by jet electrons through the charge-separation electric field.
As a result, the electron an proton density distributions are still almost identical.
While for the IGM species, they also result in almost the same density distribution due to the charge-separation field.
Note that around the jet head (where interaction is the most violent), the bow-shock structure now becomes much more complicated, due to the distortion effect of $B_{z0}$.

\begin{figure*}
	\begin{center}
	\includegraphics{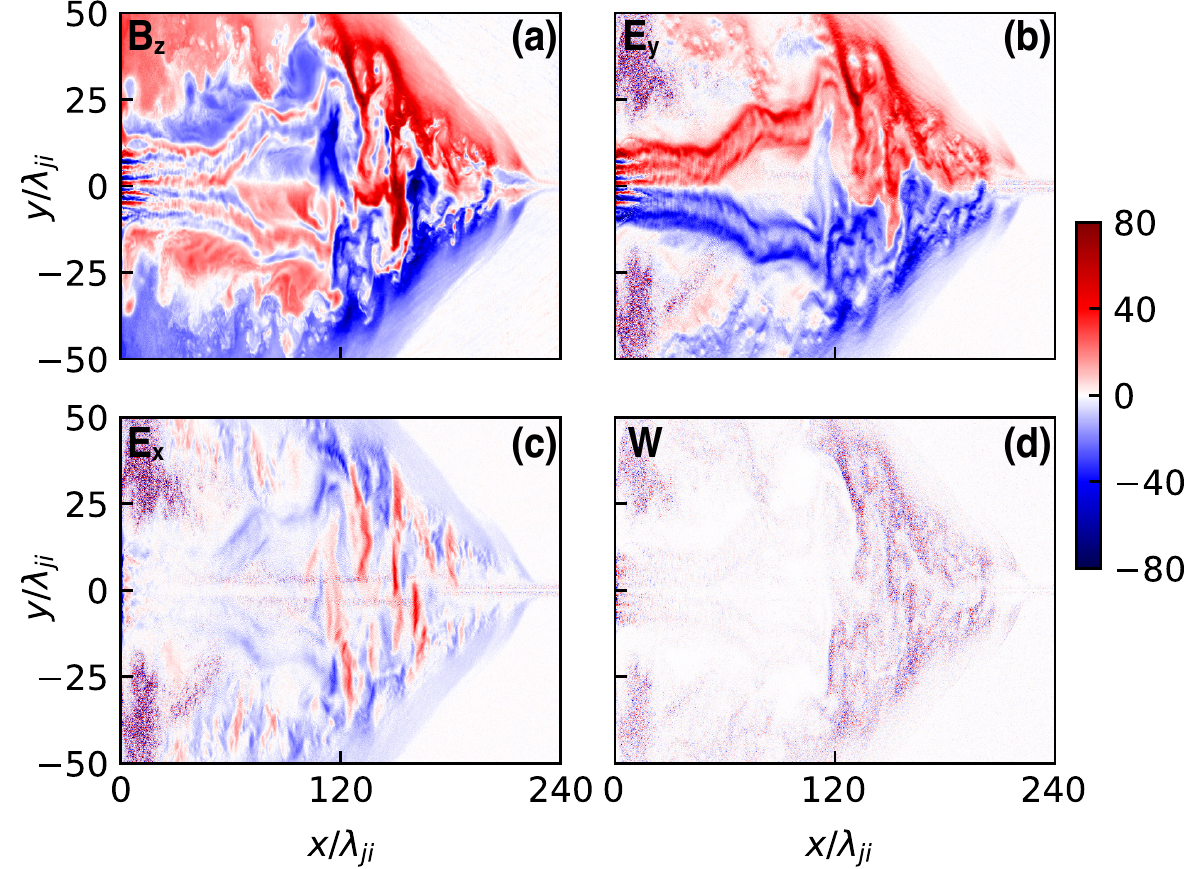}
 	\caption{(color online) EM fields ($B_z$, $E_y$ and $E_x$) and the work (done by the Lorentz force $W$) distributions at $t = 240\omega_{ji}^{-1}$ in $\sigma = 2$ case, which are also normed by $B_0$, $E_0$ and $W_0$. Note that only self-generated $B_z$ are plotted here.}
	\label{field2}
	\end{center}
\end{figure*}

To further demonstrate the distortion effect of $B_{z0}$, the EM field distributions at $t=240\omega_{ji}^{-1}$ in $\sigma = 2$ case are shown in Fig.\ref{field2}.
Specifically, around the jet body, the filaments are bifurcated in accordance with the density.
And near the jet head, the deflected filaments have been further distorted into a turbulent structure.
Besides, neither $B_z \approx E_y \gg E_x$ nor $W \approx 0$ is the case under weakly magnetized regime.
To explain, assume $E_y = E_{y1} + E_{y2}$, in which $E_{y1} \approx B_z$ is caused by the WBI. 
While $E_{y2}$ and almost all $E_x$ are from charge-separation and induction $\partial_t \bm{E} = \bm{\nabla} \times \bm{B} - \mu_0 \bm{J}$.
Because of the turbulent structure of $B_z$, its spatial gradient becomes much larger and more complex.
More importantly, with the distorted EM fields, the Lorentz force starts to do work (in Fig.\ref{field2} (d)), enabling energy exchange between particles and fields.

\begin{figure*}
	\begin{center}
	\includegraphics{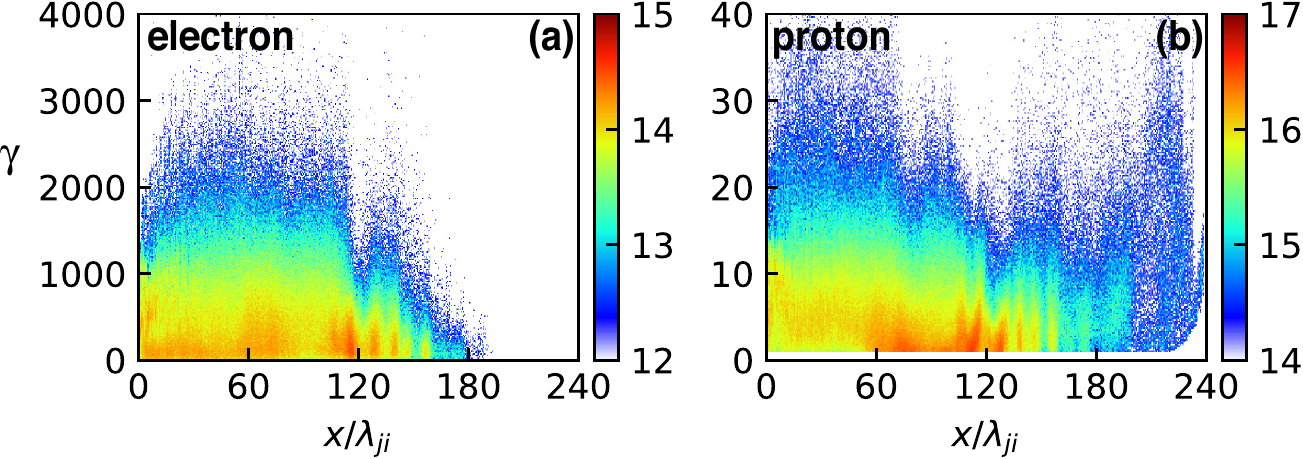}
 	\caption{(color online) Energy spatial distribution of jet electron (a) and proton (b) between $|y| < 20\lambda_{ji}$ at $t=240\omega_{ji}^{-1}$ in $\sigma = 2$ case.  The colorbar represents the particle number in logarithm.}
	\label{energy2}
	\end{center}
\end{figure*}

With the enhanced in-plane electric field in $\sigma = 2$ case, jet electrons draw much more energy from jet protons, as is shown in Fig.\ref{energy2}.
Specifically, at the injection region, although the WBI is just in linear stage, jet electrons immediately draw much energy via charge-separation field mediated by $B_{z0}$.
While at the interaction region, together with the induction electric field, jet electrons draw almost twice of energy than in $\sigma = 0$ case.
Correspondingly, jet protons lose much of their energy all along.
However, owing to the work done by the Lorentz force, many jet protons get accelerated in $\sigma = 2$ case.

In short, in $\sigma = 2$ case, jet electrons are directly diverged by magnetic fields, while jet protons are mainly dragged by electrons via the charge-separation field, which leads to much more energy transfer between them.
The density distributions of both species remain the same -- bifurcated apart at the jet body and distorted at the jet head.
Accordingly, EM fields distribute in a turbulent structure, leading to large induction in-plane electric fields.
As a result, Lorentz force starts to do work, enabling energy exchange between fields and particles, and thus proton acceleration.

\subsection{Strongly magnetized regime (for the presented case, $\sigma = 10$)}

\begin{figure*}
	\begin{center}
	\includegraphics{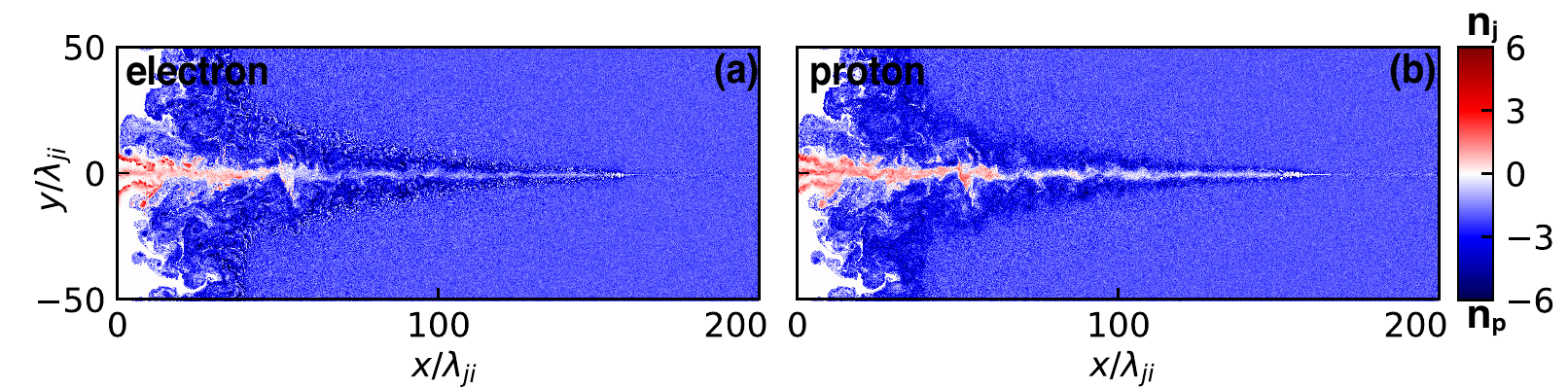}
 	\caption{(color online) Density distributions of both electron (a) and proton (b) at $t=200\omega_{ji}^{-1}$ in $\sigma = 10$ case. Jet densities are in red $n_j$, while the IGM densities are in blue $n_p$, both normalized by the initial jet number density $n_0$.}
	\label{dens2}
	\end{center}
\end{figure*}

Substituting $\sigma = 10$ into Eq. \ref{radius}, the cyclotron radius of jet protons is now $r_{ji} = 16 \lambda_{ji}$ and that of jet electrons is $r_{je} = 0.16 \lambda_{ji}$, which means jet protons should have been directly deflected by the magnetic fields, while jet electrons been almost trapped.
However, as is shown in Fig.\ref{dens2}, when considering the collective plasma effects, jet protons are directly converged by magnetic fields, but with a larger radius dragged by jet electrons through charge-separation field.
While jet electrons can still transport forward, dragged by jet protons.
Identical with each other, the density distributes in a collimated manner in $\sigma = 10$ case, which is different from the bifurcated one in $\sigma = 2$ case.

\begin{figure*}
	\begin{center}
	\includegraphics{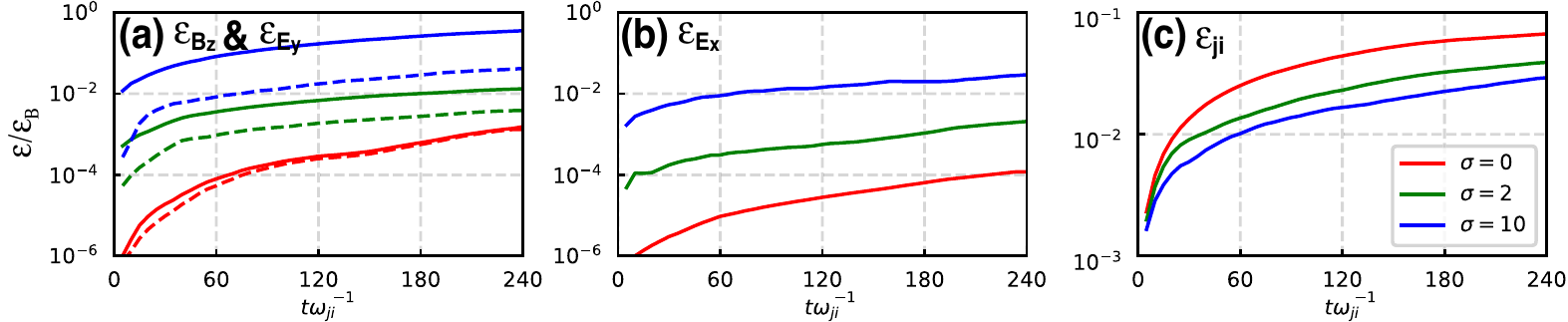}
 	\caption{(color online) Energy evolution of (a) self-generated $B_z$ (solid lines) and $E_y$ (dashed lines) (b) $E_x$ and (c) jet protons $\varepsilon_{ji}$ till the end of the transport $t=240\omega_{ji}^{-1}$. Legends are the same for all plots. All energies are normed by $\varepsilon_B$, which is the magnetic field energy at $t=240\omega_{ji}^{-1}$ in $\sigma=10$ case.}
	\label{evo}
	\end{center}
\end{figure*}

Of course, as $B_{z0}$ becomes five times larger, the charge-separation field and the induction field will also become much larger, so will the energy exchange.
Figure \ref{evo} shows the energy evolution of all three cases.
Firstly, in Fig.\ref{evo} (a), the energy of transverse EM fields becomes larger as $\sigma$ increases, so does $\varepsilon_{Bz} - \varepsilon_{Ey}$.
Because more work is done by Lorentz force as $\sigma$ increases, and particles get more energy from in-plane electric field directly. 
Secondly, in Fig.\ref{evo} (b), as $\sigma$ increases, $E_x$ energy (which contains from the charge-separation field energy and the induction field energy) becomes larger, too.
Thirdly, in Fig.\ref{evo} (c), as $\sigma$ increases, the energy exchange will become more violent, thus the jet proton energy will become smaller.

\begin{figure*}
	\begin{center}
	\includegraphics{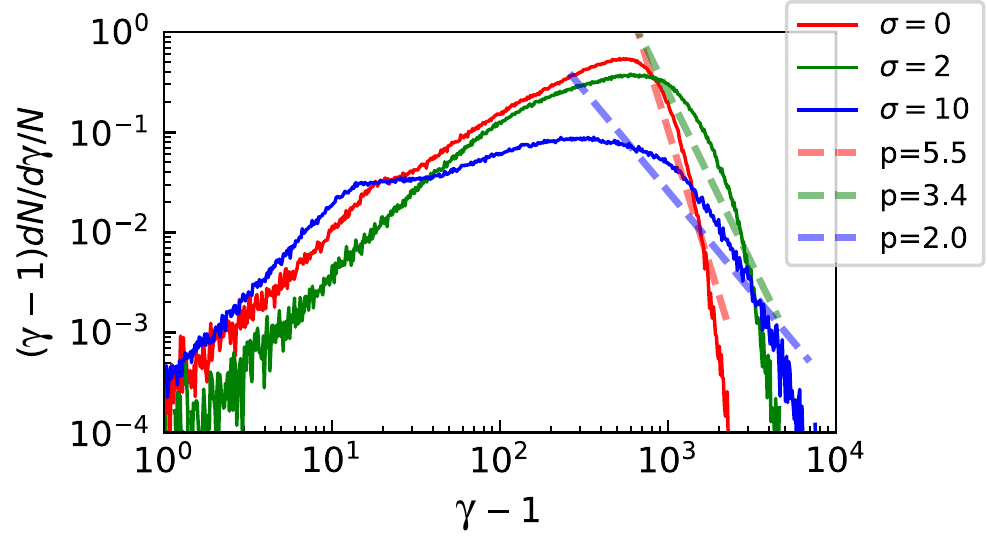}
 	\caption{(color online) SED of jet electrons for all three cases (solid lines) at $t=240\omega_{ji}^{-1}$. The power law slopes of the SED are fitted (from peak to cut-off) in dashed lines.}
	\label{sed}
	\end{center}
\end{figure*}

Last but not least, as the most important observable in astrophysics, the SED of jet electrons for all cases are compared in Fig.\ref{sed}.
Specifically, the cut-off energy increases as $\sigma$ increases, because the energy exchange is more and more violent.
Besides, the electron number gradually decreases as $\sigma$ increases, because $B_{z0}$ tends to diverge electron away from $|y|< 20\lambda_{ji}$.

More importantly, the power law slope (from each SED's peak to cut-off) is decreasing as $\sigma$ increases, i.e. $p=5.5$ when $\sigma  = 0$, $p=3.4$ when $\sigma  = 2$, and $p=2.0$ when $\sigma  = 10$.
In other words, the SED becomes flatter as the magnetic field around becomes stronger.
As $r_{s} \propto \gamma_s \sigma^{-1}$, particles with higher energy under stronger $B_{z0}$ will suffer more from the deflection, which leads to higher charge-separation field and induction field, and more violent energy exchange.
Thus, the particles in the tail of the SED tends to move to higher energy with $B_{z0}$, achieving a flatter spectrum with a smaller slope.
It is worth mention that the observed SEDs of AGN jets are also characterized by flat electron spectra with the power law slope $p < 2$ \cite{Sikora2009, Sironi2011}.

\section{CONCLUSION}

In conclusion, we study the kinetic effects of perpendicular magnetic fields on the transport of relativistic jet in the IGM. 
Through 2D PIC simulations, we find the perpendicular magnetic field tends to pull different charged particles apart, leading to charge-separation fields via collective plasma effects.
Specifically, when the IGM is weakly magnetized, electrons are directly diverged by magnetic fields, while protons are mainly dragged by electrons via charge-separation field, resulting in identical bifurcation distribution.
However, when the IGM is strongly magnetized, the contrary is the case.
In both regimes, the balance between magnetic field deflection and charge-separation field drag tremendously distorts the jet density and EM fields distributions.
As a result, the enlarged in-plane electric fields bring about much more energy exchange from protons to electrons, especially for the most energetic ones. 
Because the particles in the tail of the SED tend to move to higher energy, flatter SED with smaller power law slopes will be achieved.
Comparing different cases, we find that the power law slope of the SED decreases as the magnetization rate increases.
%Besides, the spatial gradient of magnetic fields leads to the induction electric field, through which the Lorentz force starts to do work, enabling the energy transfer between fields and particles.

With the kinetic simulation revealing the microscopic particle dynamics, this detailed study of the magnetized jet-IGM interaction may offer us potential explanations about the formation of a particular SED in astrophysics. 

\section{ACKNOWLEDGEMENT}

This work is supported by the NSAF, Grant No.U1630246 and No.U1730449; the National Key Program of S\&T Research and Development, Grant No.2016YFA0401100; the National Natural Science Foundation of China, Grants Nos.11575298, 11575011, 91230205, 11575031, and 11175026; the National Science Challenge Project TZ 2016005; the National Basic Research 973 Projects No.2013CBA01500 and No.2013CB834100, and the National High-Tech 863 Project. B.Q. acknowledges the support from Thousand Young Talents Program of China. The computational resources are supported by the Special Program for Applied Research on Super Computation of the NSFC-Guangdong Joint Fund (the second phase) under Grant No.U1501501. Y. W. P would like to thank Z. Z. Q for valuable discussions about astrophysical backgrounds.

\section*{REFERENCES}

\bibliography{mybib}{}
\bibliographystyle{iopart-num}

\end{document}